\documentclass[preprint,12pt]{elsarticle}

\usepackage{graphicx}
\usepackage{amssymb}
\usepackage{subfigure}
\usepackage{lineno}
\usepackage[load-configurations = abbreviations]{siunitx}
\usepackage{xspace}
\usepackage{textcomp}
\usepackage{float}

\newcommand{\VGL}{$\mathrm{V}_{\mathrm{GL}}$\xspace}
\newcommand{\VFD}{$\mathrm{V}_{\mathrm{FD}}$\xspace}
\newcommand{\VBD}{$\mathrm{V}_{\mathrm{BD}}$\xspace}

\journal{NIMA}

\begin{document}

\begin{frontmatter}


\title{Layout and Performance of HPK Prototype LGAD Sensors for the High-Granularity Timing Detector}

\author[e]{X. Yang}
\author[i]{S. Alderweireldt}
\author[f]{N. Atanov}
\author[a]{M.K. Ayoub}
\author[a]{J. Barreiro Guimarães da Costa}
\author[b]{L. Castillo Garc\'ia}
\author[e]{H. Chen}
\author[h]{S. Christie}
\author[j]{V. Cindro}
\author[a,k]{H. Cui}
\author[d]{G. D'Amen}
\author[f]{Y. Davydov}
\author[a]{Y.Y. Fan}
\author[h]{Z. Galloway}
\author[e]{J.J. Ge}
\author[h]{C. Gee}
\author[d]{G. Giacomini}
\author[b]{E.L. Gkougkousis}
\author[b]{C. Grieco}
\author[b]{S. Grinstein}
\author[g]{J. Grosse-Knetter}
\author[i]{S. Guindon}
\author[a,k]{S. Han}
\author[j]{A. Howard}
\author[a]{Y.P. Huang}
\author[h]{Y. Jin}
\author[a,k]{M.Q. Jing}
\author[a]{R. Kiuchi}
\author[j]{G. Kramberger}
\author[i]{E. Kuwertz}
\author[h]{C. Labitan}
\author[g]{J. Lange}
\author[c]{M. Leite} 
\author[e]{C.H. Li}
\author[e]{Q.Y. Li}
\author[a]{B. Liu}
\author[a]{J.Y. Liu}
\author[e]{Y.W. Liu}
\author[e]{H. Liang}
\author[a]{Z.J. Liang}
\author[h]{M. Lockerby}
\author[a]{F. Lyu}
\author[j]{I. Mandić}
\author[h]{F. Martinez-Mckinney}
\author[h]{S.M. Mazza}
\author[j]{M. Mikuž}
\author[h]{R. Padilla} 
\author[a,k]{B.H. Qi}
\author[g]{A. Quadt}
\author[a,k]{K.L. Ran}
\author[h]{H. Ren}
\author[i]{C. Rizzi}
\author[d]{E. Rossi}
\author[h]{H.F.-W. Sadrozinski}
\author[c]{G.T. Saito}
\author[h]{B. Schumm}
\author[g]{M. Schwickardi}
\author[h]{A. Seiden}
\author[a]{L.Y. Shan}
\author[a]{L.S. Shi}
\author[a]{X. Shi}
\author[i]{A. Soares Canas Ferreira}
\author[e]{Y.J. Sun}
\author[a,k]{Y.H. Tan}
\author[d]{A. Tricoli}
\author[e]{G.Y. Wan}
\author[h]{M. Wilder} 
\author[a,k]{K.W. Wu}
\author[h]{W. Wyatt} 
\author[a,k]{S.Y. Xiao}
\author[a,k]{T. Yang}
\author[a]{Y.Z. Yang}
\author[a,k]{C.J. Yu}
\author[e]{L. Zhao}
\author[a]{M. Zhao}
\author[h]{Y. Zhao}
\author[e]{Z.G. Zhao}
\author[e]{X.X. Zheng}
\author[a]{X.A. Zhuang}

\address[b]{Institut de F\'isica d'Altes Energies (IFAE) Carrer Can Magrans s/n, Edifici Cn, Universitat Aut\'onoma de Barcelona (UAB), E-08193 Bellaterra (Barcelona), Spain }

\address[k]{University of Chinese Academy of Sciences, 19A Yuquan Road, Shijingshan District, Beijing 100049, China}

\address[a]{Institute of High Energy Physics, Chinese Academy of Sciences, 19B Yuquan Road, Shijingshan District, Beijing 100049, China}

\address[f]{Joint Institute for Nuclear Research, Joliot-Curie street 6, Dubna, 141980 Russia	}

\address[i]{CERN, Esplanade des Particules 1, 1211 Geneva 23}
\address[g]{II. Physikalisches Institut, Georg-August-Universit\"{a}t G\"ottingen, Friedrich-Hund-Platz 1, 37077 G\"ottingen, Germany}
\address[e]{Department of Modern Physics and State Key Laboratory of Particle Detection and Electronics, University of Science and Technology of China, Hefei 230026, China}

\address[j]{Jozef Stefan Institut (JSI), Dept. F9, Jamova 39, SI-1000 Ljubljana, Slovenia} 
\address[h]{SCIPP, Univ. of California Santa Cruz, CA 95064, USA}

\address[c]{Instituto de F\'isica, Universidade de S\~ao Paulo, S\~ao Paulo (USP), R. do Mat\~ao, 1371, Cidade Universit\'aria, S\~ao Paulo - SP 05508-090 - Brazil}

\address[d]{Brookhaven National Laboratory (BNL), Upton, NY 11973, U.S.A.}

\begin{abstract}
The High-Granularity Timing Detector is a detector proposed for the ATLAS Phase II upgrade. The detector, based on the Low-Gain Avalanche Detector~(LGAD) technology will cover the pseudo-rapidity region of $2.4<|\eta|<4.0$ with two end caps on each side and a total area of \SI{6.4}{m^2}. The timing performance can be improved by implanting an internal gain layer that can produce signal with a fast rising edge, which improve significantly the signal-to-noise ratio. The required average timing resolution per track for a minimum-ionising particle is \SI{30}{ps} at the start and \SI{50}{ps} at the end of the HL-LHC operation. This is achieved with several layers of LGAD. The innermost region of the detector would accumulate a \SI{1}{M\eV}-neutron equivalent fluence up to \SI{2.5e15}{\per\centi\metre\squared} before being replaced during the scheduled shutdowns. The addition of this new detector is expected to play an important role in the mitigation of high pile-up at the HL-LHC. The layout and performance of the various versions of LGAD prototypes produced by Hamamatsu~(HPK) have been studied by the ATLAS Collaboration. The breakdown voltages, depletion voltages, inter-pad gaps, collected charge as well as the time resolution have been measured and the production yield of large size sensors has been evaluated.
\end{abstract}


\end{frontmatter}


\section{Introduction}
\label{S:1}

Low-Gain Avalanche Detectors (LGAD)~\cite{SADROZINSKI201618} have become one of the most precise detector technologies for timing measurements due to their internal gain mechanism. It was proposed by the RD50 collaboration and developing by CNM~\cite{pellegrini2014technology}, FBK~\cite{fbkart}, BNL~\cite{giacomini2019development} and HPK~\cite{GALLOWAY201919}. Recent developments concentrate on the optimization of the pad periphery and the doping profile to prevent premature breakdown and maintain the performance of the sensor after extremely high fluences of hadron irradiation. After several years of research, the properties of LGADs have become better known. In order to apply the technology in the ATLAS HGTD upgrade~\cite{TPCollaboration:2623663}, the sensors need to maintain a time resolution per hit better than \SI{70}{\pico\second} after \SI{2.5e15}{\per\centi\metre\squared} \SI{1}{M\eV}-neutron equivalent fluence. HPK has produced the first batch of wafers of LGADs to HGTD specifications and these sensors have been tested by the ATLAS HGTD sensor group with methods including I-V (Current-voltage) curves, C-V (Capacitance–voltage) profiling, TCT~(Transient Current Technique)~\cite{mschLGAD}, beta scope~\cite{ZHAO2019387}, and several test beams~\cite{Allaire_2018}. The performances of these sensors before irradiation are shown in this paper.

\begin{figure}[htp]
\centering
\includegraphics[width=0.55\linewidth]{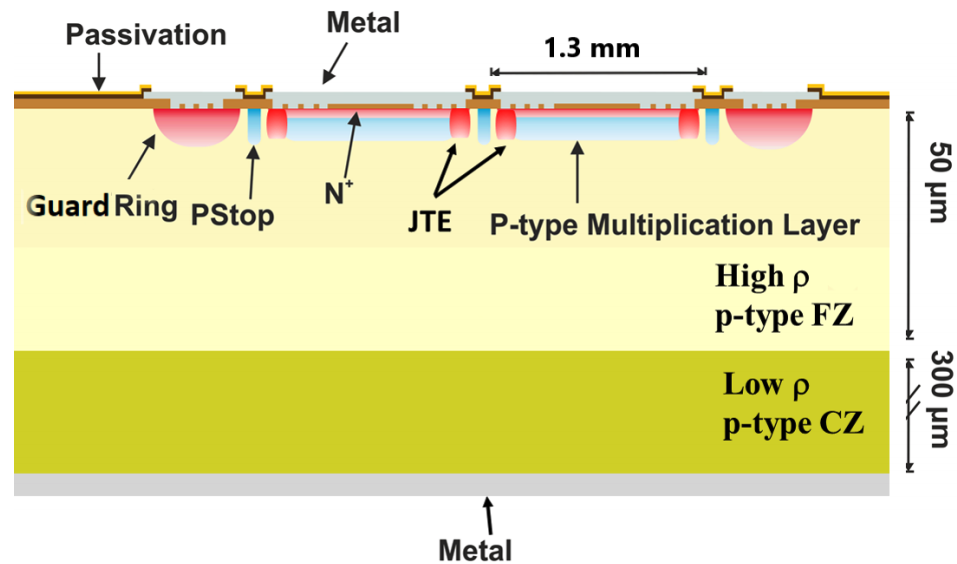}
\caption{Cross section of a sensor with \num{2 x 2} array of pads and \SI{50}{\micro\metre} active thickness~\cite{RD50CNMSensor}. }
\label{Fig:1}
\end{figure}

With a common mask, sensors with an active thickness of \SI{35}{\micro\metre} (HPK-1.1-35, HPK-1.2-35 and HPK-2-35) and \SI{50}{\micro\metre} (HPK-3.1-50 and HPK-3.2-50) were produced by HPK. The different types refer to different combinations of doping profile of the gain layer and resistivity of the bulk. The cross section of a sensor with a \num{2 x 2} array of pads is shown in Figure~\ref{Fig:1}. The layout of the arrays is based on the \SI{1.3 x 1.3}{\milli\metre} pad size. 

Single pads with different distances to the edge and \num{2 x 2} arrays with different inter-pad (IP) gap were fabricated to determine a safe separation between pads. Large arrays with \num{5 x 5} and \num{15 x 15} pads were produced and studied for the first time to demonstrate the feasibility of large sensors and to investigate the yield, for which the sensors are tested and classified in terms of breakdown voltages. The layouts of the HPK sensors are presented in Table~\ref{Tab:1}. The dynamic properties are measured with charged particles or infrared laser. The collected charge and time resolutions are measured at different voltages. 

LGADs with \SI{35}{\micro\metre} active thickness have higher capacitance, smaller generated charge and larger power dissipation after irradiation than those with \SI{50}{\micro\metre} active thickness~\cite{Xintalk}. Hence, it has been decided that \SI{35}{\micro\metre} sensors will not be used for the HGTD project. The detailed studies including IP gap, yield and uniformity only focus on the HPK-3.1-50 and HPK-3.2-50 sensors. 

\begin{table}[htp]
    \begin{center}
        \begin{tabular}{|c|c|c|c|}
            \hline
            Size &Edge [\si{\micro\metre}]&Nominal IP Gap [\si{\micro\metre}]&Total \# per wafer\\
            \hline
            Single& 200,300,500    & - &132 \\     
            \hline
            \num{2 x 2} arrays& 200,300,500    & 30,50,70,95 &70  \\ 
            \hline
            \num{3 x 3} arrays& 300    & 95  &4 \\ 
            \hline
            \num{5 x 5} arrays& 500    &95  &18 \\ 
            \hline 
            \num{15 x 15} arrays& 500    &95  &8 \\ 
            \hline   
        \end{tabular}
    \end{center} 
    \caption{Layout of the HPK Prototype sensors.}
    \label{Tab:1}
\end{table}

\begin{figure}[htp]
    \begin{center}
        \includegraphics[width=0.6\linewidth]{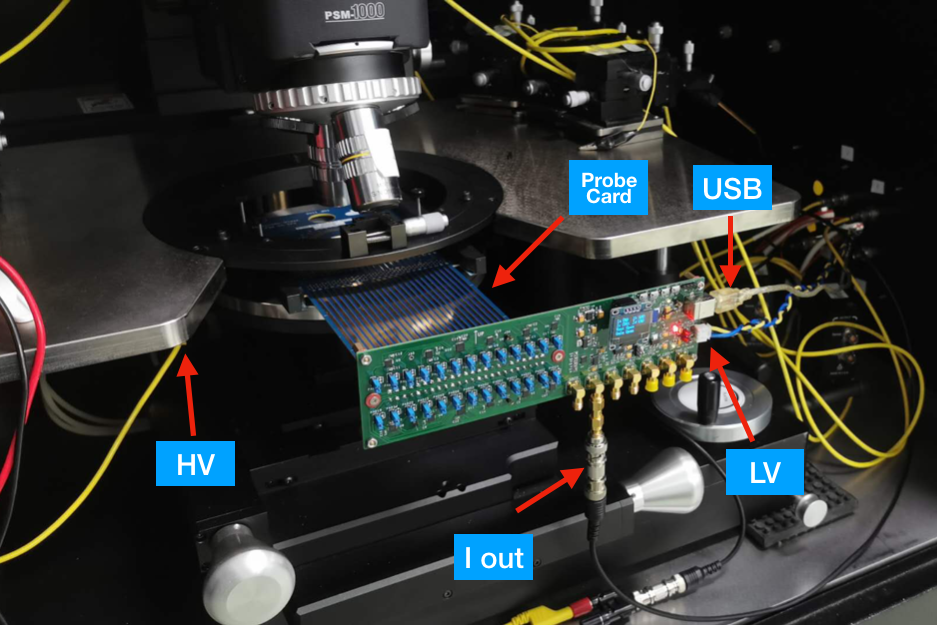}
        \caption{Setup of the HPK \num{5 x 5} pad array measurement with a dedicated probe card and a digital channel switch board. The channel under measurement is selected (I out) by a channel switch board which is controlled by a PC via the USB connection. High voltage (HV) to bias the sensor reversely is supplied by an SMU from the metal chuck contacted to the backside of the sensor. Up to four channels could be selected out by the switch board to speed up the measurement.}
        \label{Fig:11}
    \end{center}
\end{figure}

\section{Setup for I-V and C-V measurements}
The I-V and C-V measurements are proceed under probe stations. The probe needles are used for sensors with single pad and \num{2 x 2} pad array for which up to five needles are used to make the guard ring or neighbor pads grounded. For \num{5 x 5} pad array, a dedicated probe card is designed to contact all 25 pads and guard ring simultaneously as shown in Figure~\ref{Fig:11}. The channel under measurement is selected by a channel switch board while other channels remain grounded. The switch board could be controlled by computer via the USB connection so that a scan of total 25 pad could be performed semi-automatically. For \num{15 x 15} pad array, the measurements are taking under a automatic probe station with the probe needle, the guard ring is grounded while other pads not measured remain floating. The probe card for \num{15 x 15} sensors is under developing and would be applied in the future measurement. 

The high reverse bias voltage (HV) is supplied by the metal chuck of the probe station. All sensors before irradiation are measured under room temperature (about \SI{25}{^\circ C}) and relative humidity around 20\%-50\% depending on the weather. The sensors are protected from light by the black shield box during the measurement. 

For the I-V measurement, the sensor is connected with source measure unit (SMU) which can bias sensors up to around \SI{800}{\volt} and measure the current at sub-\si{\pico\ampere} level. The high precision current meter with \SI{1}{\femto\ampere} precision is series connected to the channel under test when a high precision I-V is needed. The compliance current is set to several \si{\micro\ampere} to avoid the potential damage to the instruments or sensors when breakdown. A time interval large than \SI{500}{\milli\second} and a ramping speed lower than \SI{100}{\milli\volt\per\volt} are assured at each bias point to ensure stability of the measurement circuit. 

As for C-V measurement, besides the SMU to provide the HV, a LCR meter is connected in parallel by an HV adaptor~\cite{agilent2009impedance} to measure the AC character of the device. Apart form the current compliance set in I-V,  a ziener diode is connected with sensor in parallel to limit the voltage in a safe range for LCR. The "Cp-Rp" working mode and a \SI{10}{\kilo\hertz} frequency is used for un-irradiated sensors as the recommendation of RD50~\cite{RD50Co}. An AC amplitude of \SI{100}{\milli\volt} and a step of \SI{0.25}{\volt} is used to measure the character of the gain layer. The step is further limited to \SI{0.1}{\volt} near the depletion voltage to have the fine structure measured. In order to ensure stability, the capacitance is read out every \SI{500}{\milli\second}, after the bias voltage is set, and is saved only when the relative difference with respect to the previous measurement is less than 0.1\%. 


\begin{figure}[htp]
    \begin{center}
        \includegraphics[width=0.6\linewidth]{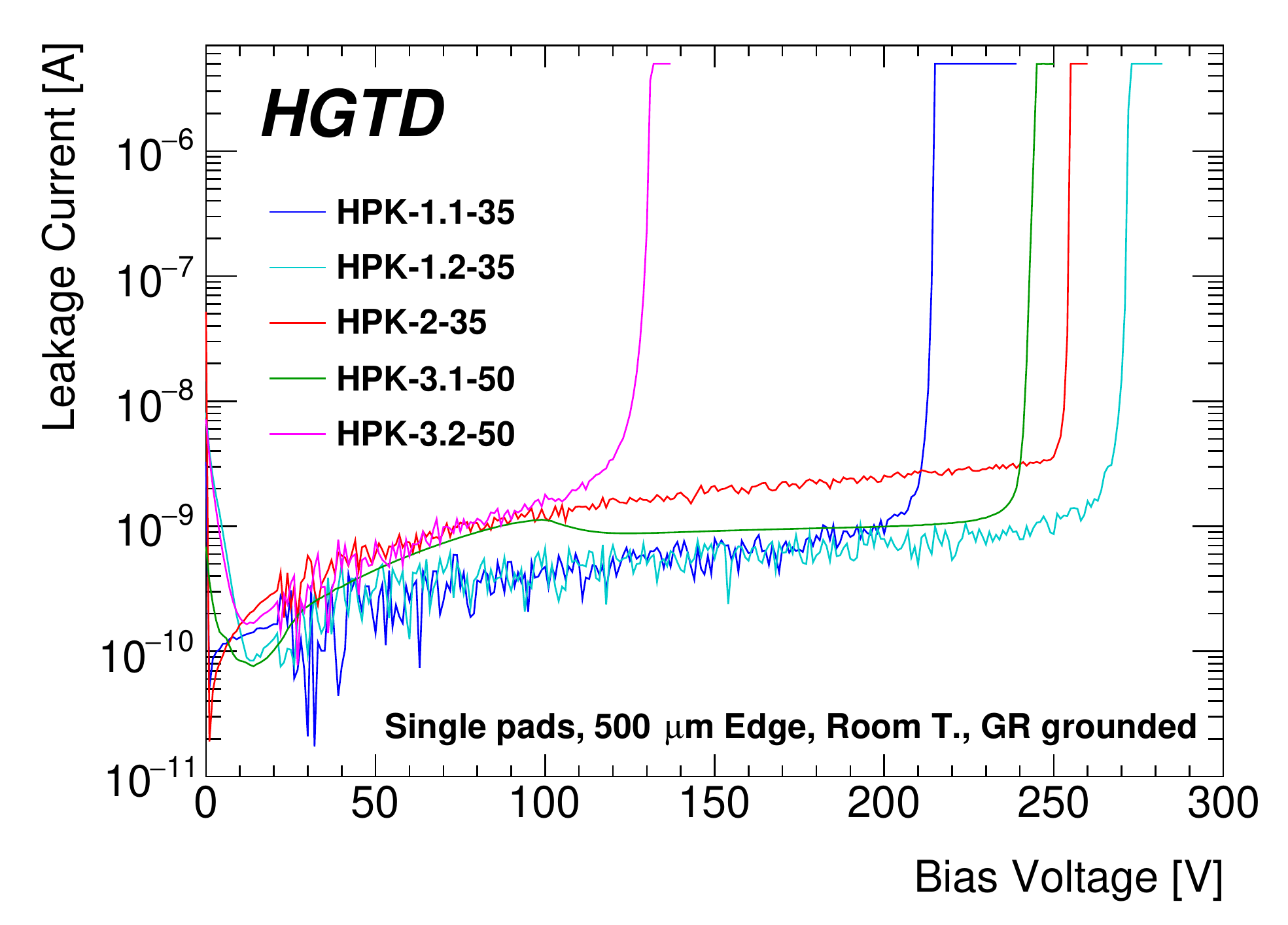}
        \caption{I-V curves of different types of HPK prototype sensor. The measurements are taken by the single needle with guard rings grounded.}
        \label{Fig:2}
    \end{center}
\end{figure}

\section{I-V, uniformity and yield}
\label{S:3}
\begin{figure}[htp]
    \begin{center}
        \subfigure[]{   \begin{minipage}[t]{0.5\linewidth}  \includegraphics[width=1.0\linewidth]{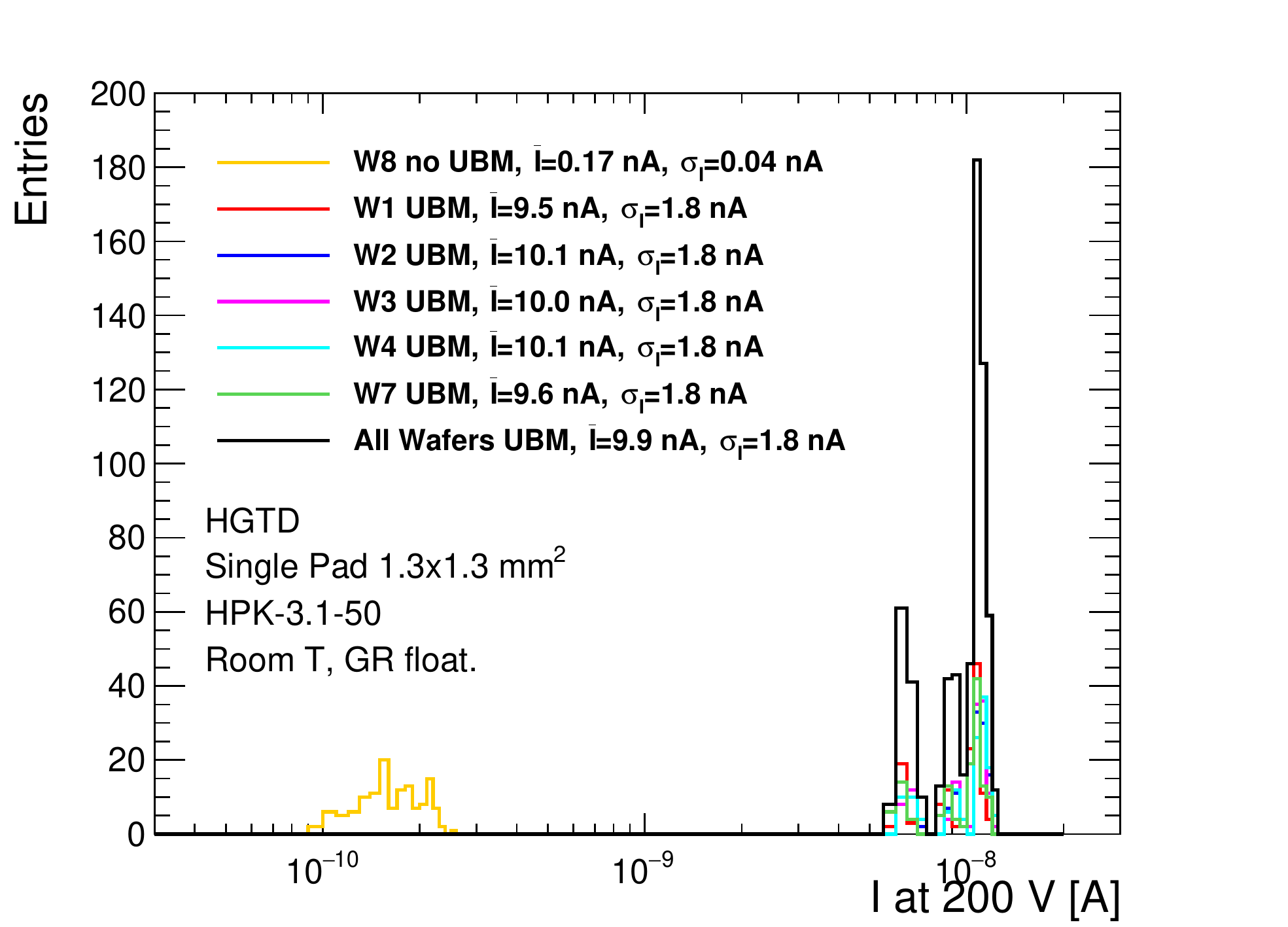}  \end{minipage}   }%
        \subfigure[]{   \begin{minipage}[t]{0.5\linewidth} \includegraphics[width=1.0\linewidth]{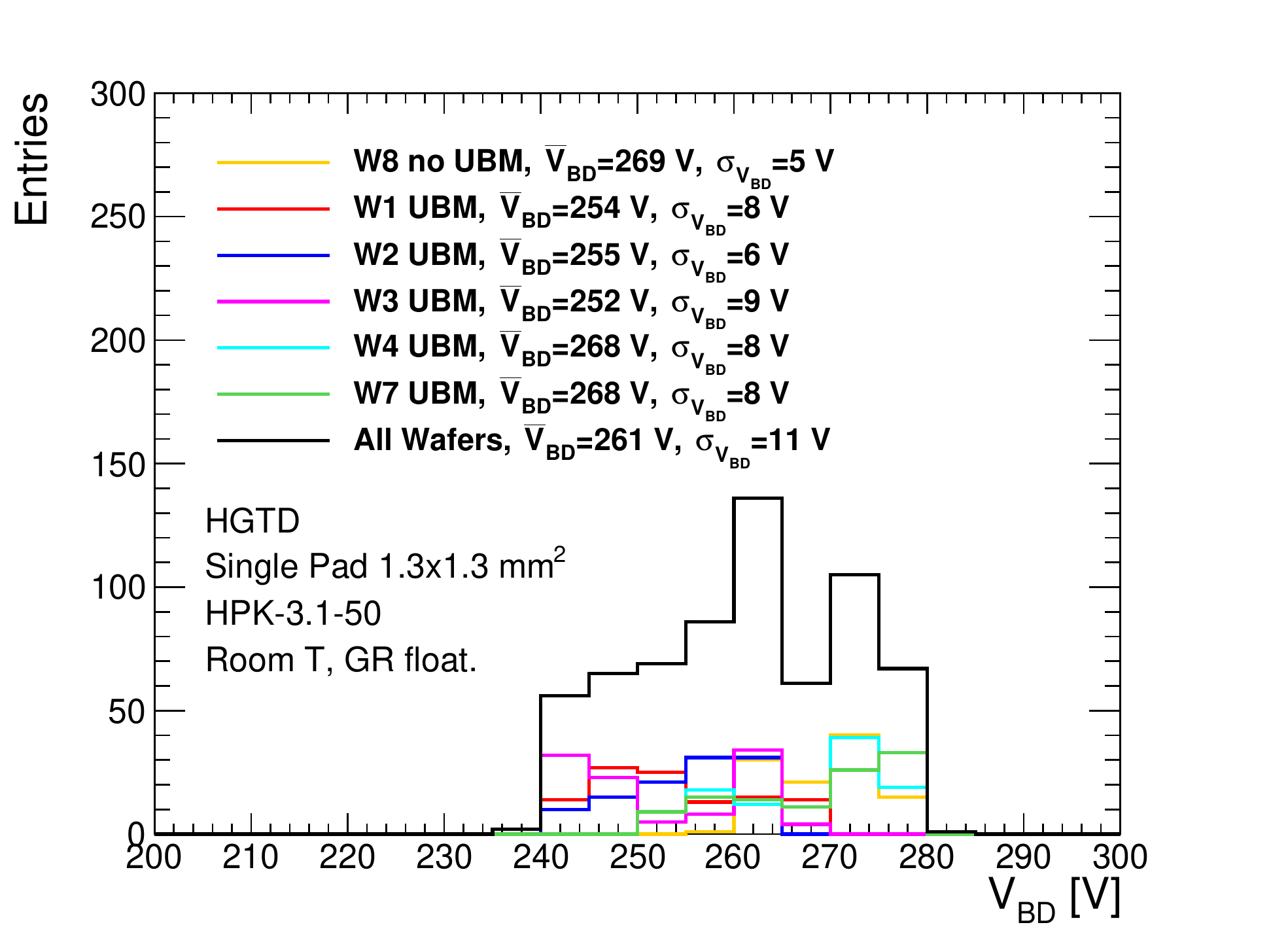} \end{minipage}  }%
        \vskip\baselineskip
        \subfigure[]{   \begin{minipage}[t]{0.5\linewidth}  \includegraphics[width=1.0\linewidth]{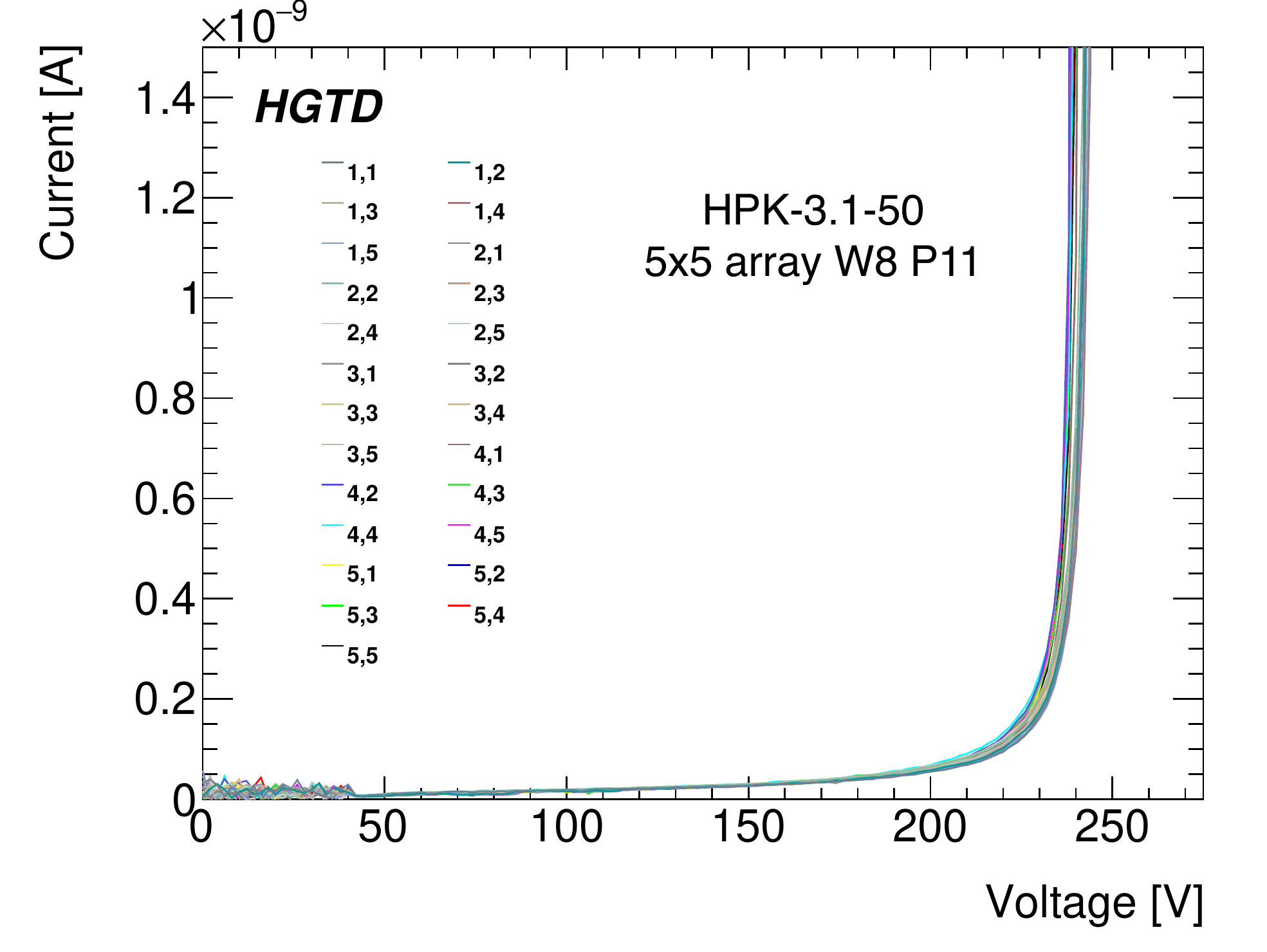}  \end{minipage} }%
        \subfigure[]{   \begin{minipage}[t]{0.5\linewidth}  \includegraphics[width=1.0\linewidth]{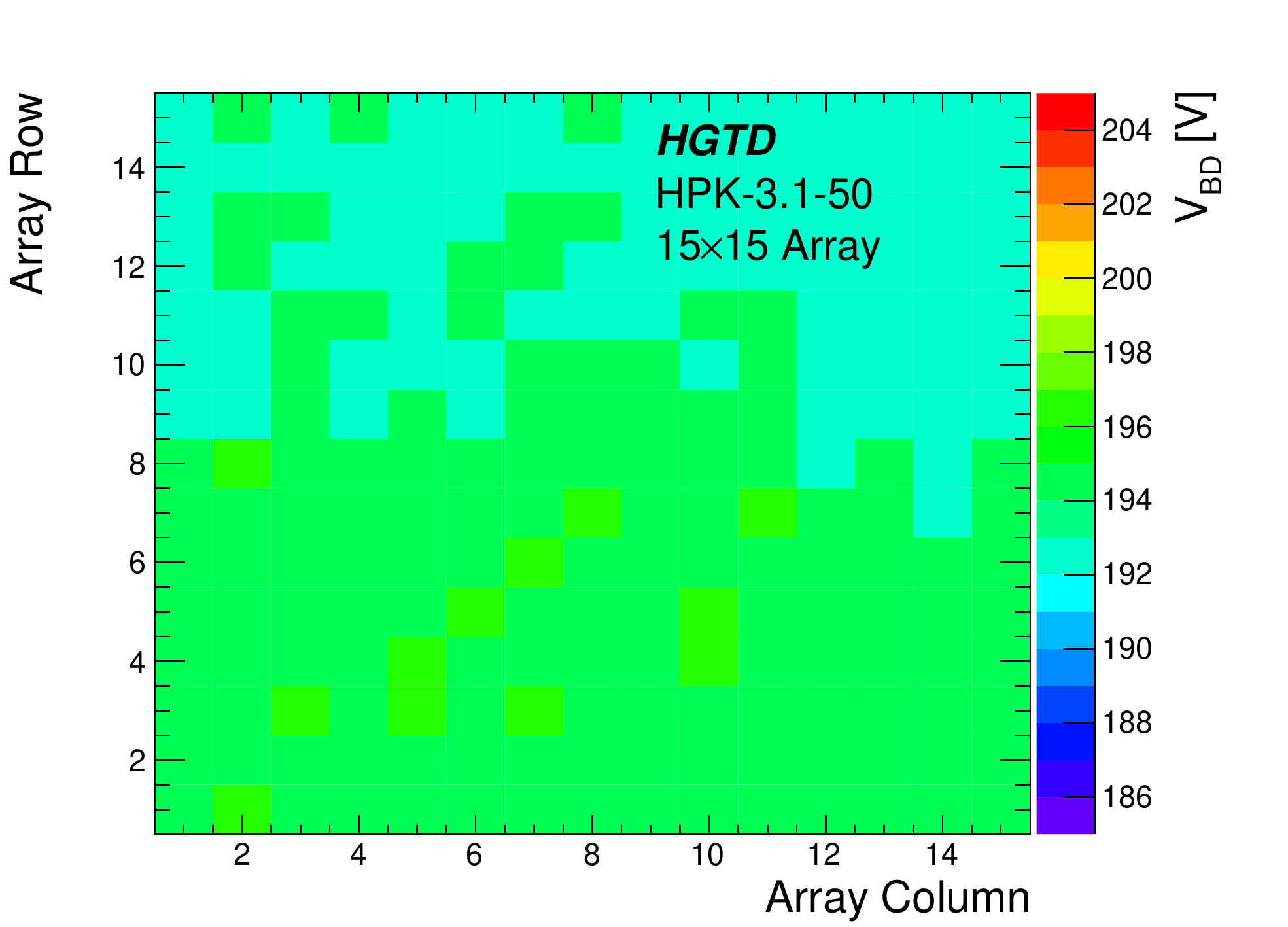}\end{minipage} }%
    \end{center}
    \caption{Figures to show the uniformity of the HPK-3.1-50 sensors~\cite{HGTDPublicPlots}. (a) Distribution of LGAD leakage current at \SI{200}{V}. (b) Distribution of breakdown voltages. (c) The I-V curves of the 25 pads in an HPK-3.1-50 \num{5 x 5} array.  (d) Distribution of pad breakdown voltages of a \num{15 x 15} array~\cite{mschLGAD}. (a) and (b) are measured with automatic probe station and guard ring floating. (c) is measured by probe card with all pads and guard ring grounded. (d) is taken by a semi-automatic probe station with guard ring and other pads floating. }
    \label{Fig:3}
\end{figure}

The I-V for each single pad sensor is measured at room temperature with single needle and the guard ring grounded. The results are shown in Figure~\ref{Fig:2}.  The distribution of leakage current and breakdown voltage on HPK-3.1-50 single pad sensors are shown in Figure~\ref{Fig:3}(a) and (b). In Figure~\ref{Fig:3}(c), the I-V curves of each pad on an HPK-3.1-50 \num{5 x 5} sensor measured with a probe card show a good uniformity. The breakdown voltage (\VBD) is defined with these curves by the bias voltage corresponding to a leakage current of \SI{50}{\nano\ampere}.

The performances of different types of sensors are compared on single pads with exactly the same geometry. For the \SI{35}{\micro\metre} sensors, the HPK-1.1-35 sensors have breakdown voltages of \SI{215}{V}, while type HPK-1.2-35 with the similar gain layer implantation as type HPK-1.1-35 but a higher bulk resistivity shows a higher breakdown voltage of \SI{270}{V}. The HPK-2-35 sensor which has the similar bulk resistivity as HPK-1.1-35 but shallower and wider doping shows a higher breakdown voltage of \SI{254}{V}. For the \SI{50}{\micro\metre} sensors, the major difference between HPK-3.1-50 and HPK-3.2-50 is the depth of the gain layer implantation. For HPK-3.1-50 which has a shallower implantation, the breakdown voltage is \SI{242}{V}. HPK-3.2-50 has a deeper implantation and its breakdown voltage is \SI{130}{V}, which is much lower. All sensors are measured to have a low leakage current level (sub-\si{\nano\ampere}) before breakdown. 

\begin{figure}[htp]
    \begin{center}
        \includegraphics[width=0.55\linewidth]{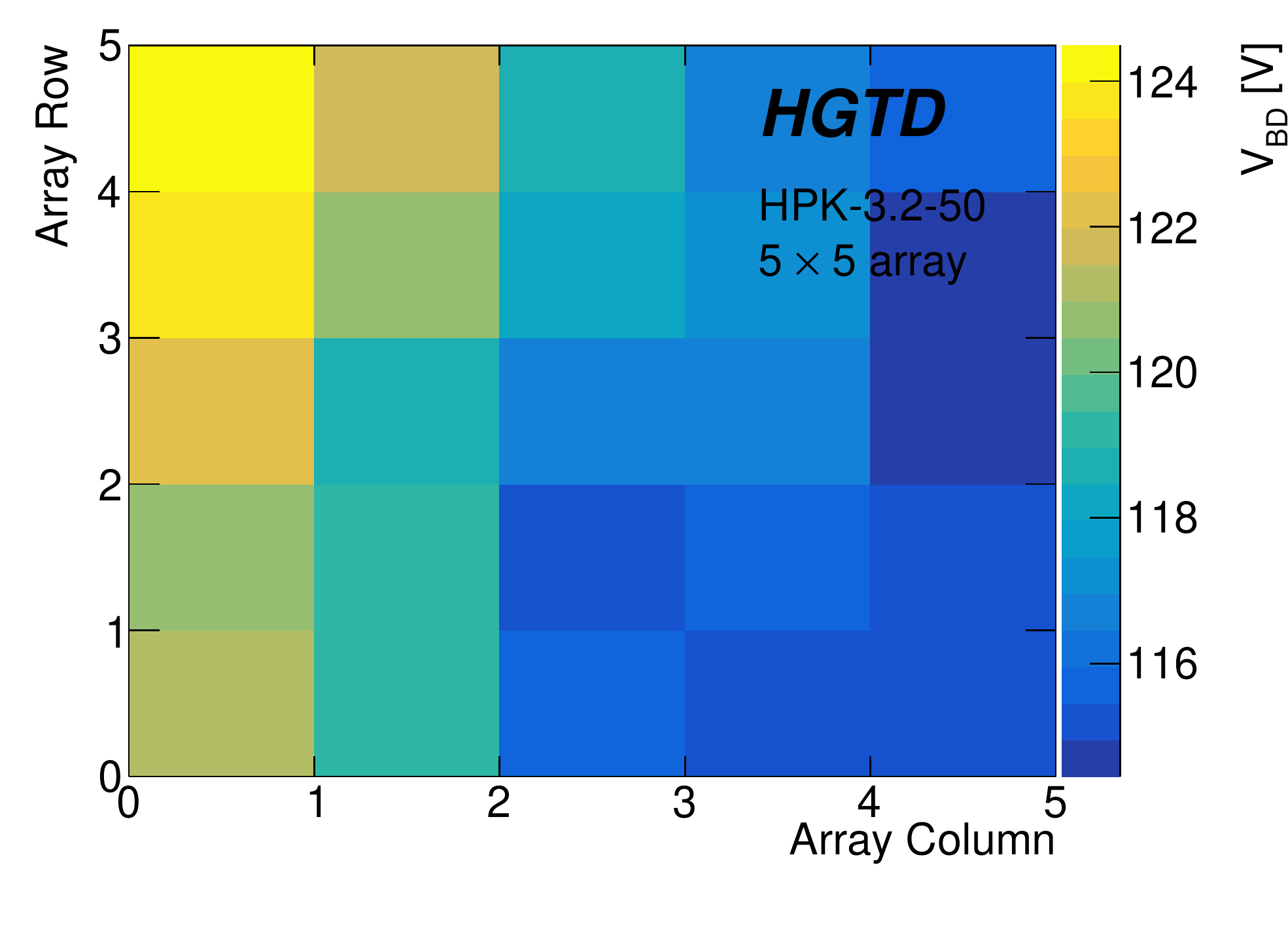}
    \end{center}
    \caption{Distribution of \VBD on an HPK-3.2-50 \num{5 x 5} pads array. The measurement is taken by a dedicate probe card with guard ring and neighbor pads grounded.}
    \label{Fig:5}
\end{figure}

Since the HGTD needs a very large number of modules with \num{30 x 15} arrays of LGADs, the large scale yield is important for the sensor fabrication. We have summarized the fraction of good pads and perfect sensors in different array size for the yield estimation (see Table~\ref{Tab:2}). The good pads are defined as pads having a breakdown voltage above 90\% of the nominal value and a perfect sensor is a sensor with 100\% good pads in it. The fraction of perfect sensors is 100\% for all sensors with single pad or \num{2 x 2}, \num{5 x 5} arrays of pads. For sensors with \num{15 x 15} arrays of pads, the fraction of perfect sensors is 85.2\% for HPK-3.1-50, 91.3\% for HPK-3.2-50 and the fraction of good pads is 99.5\% for HPK-3.1-50, 99.8\% for HPK-3.2-50. 

The uniformity of the pads in large arrays is also important. For all HPK-3.1-50 wafers evaluated, the uniformity has been shown by the distributions of leakage current, breakdown voltage and the breakdown voltage of all pads in a \num{15 x 15} array in Figure~\ref{Fig:3}. The distribution of breakdown voltage on an HPK-3.2-50 \num{5 x 5} array is shown in Figure~\ref{Fig:5}. An acceptable uniformity has been observed. It is also observed that the breakdown voltages of arrays are slightly lower than those of single pad sensors, most likely due to the differences in surrounding structures.


\begin{table}[htp]
\begin{center}
\includegraphics[width=0.8\linewidth]{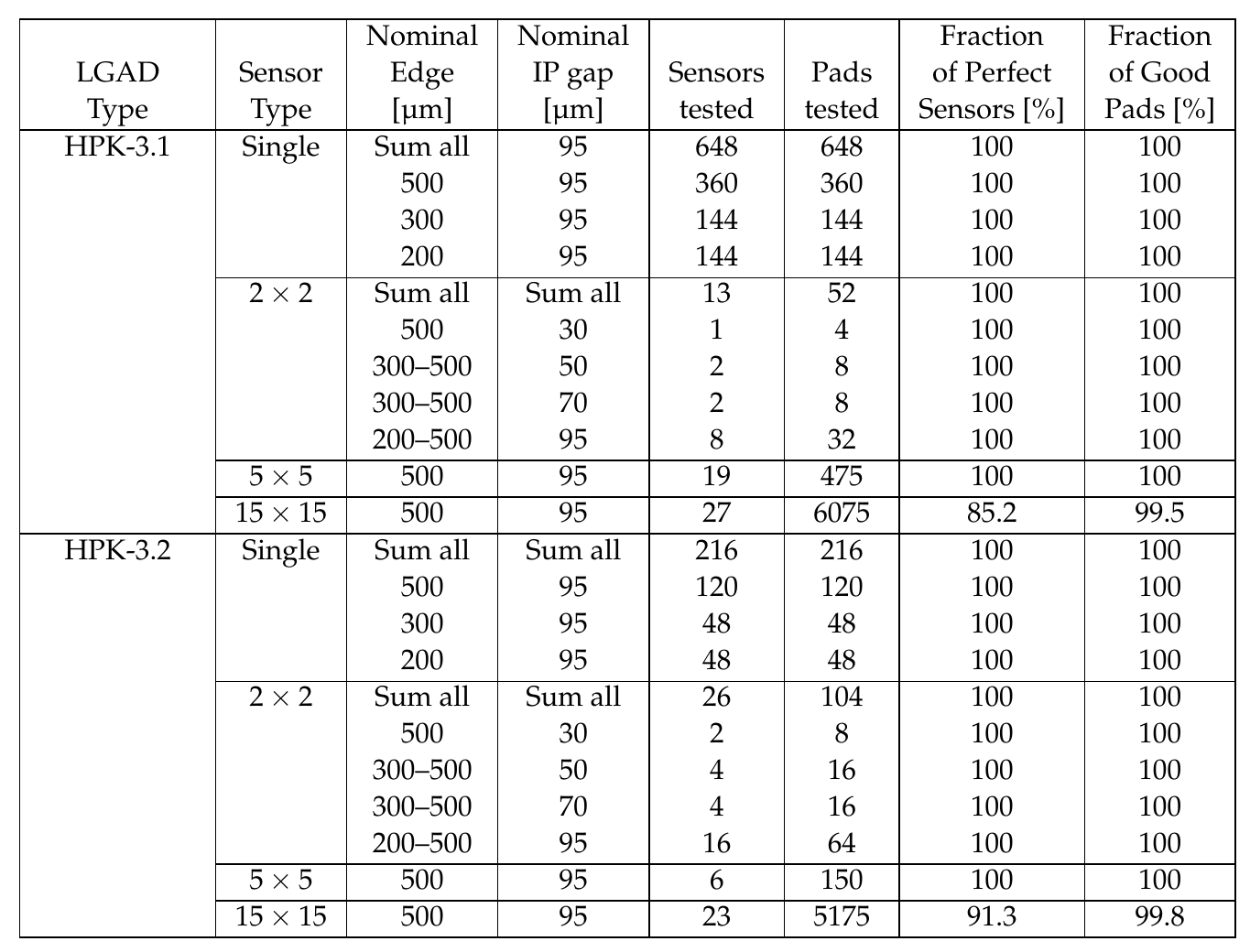}
\end{center}
\caption{Large scale yield table estimated from the first HPK prototype production.}
\label{Tab:2}
\end{table}

\section{C-V and depletion voltage}
\label{S:4}

\subsection{Single pad C-V of each type}
\begin{figure}[htp]
\begin{center}
\subfigure[]{
\begin{minipage}[t]{0.55\linewidth}
\includegraphics[width=1.0\linewidth]{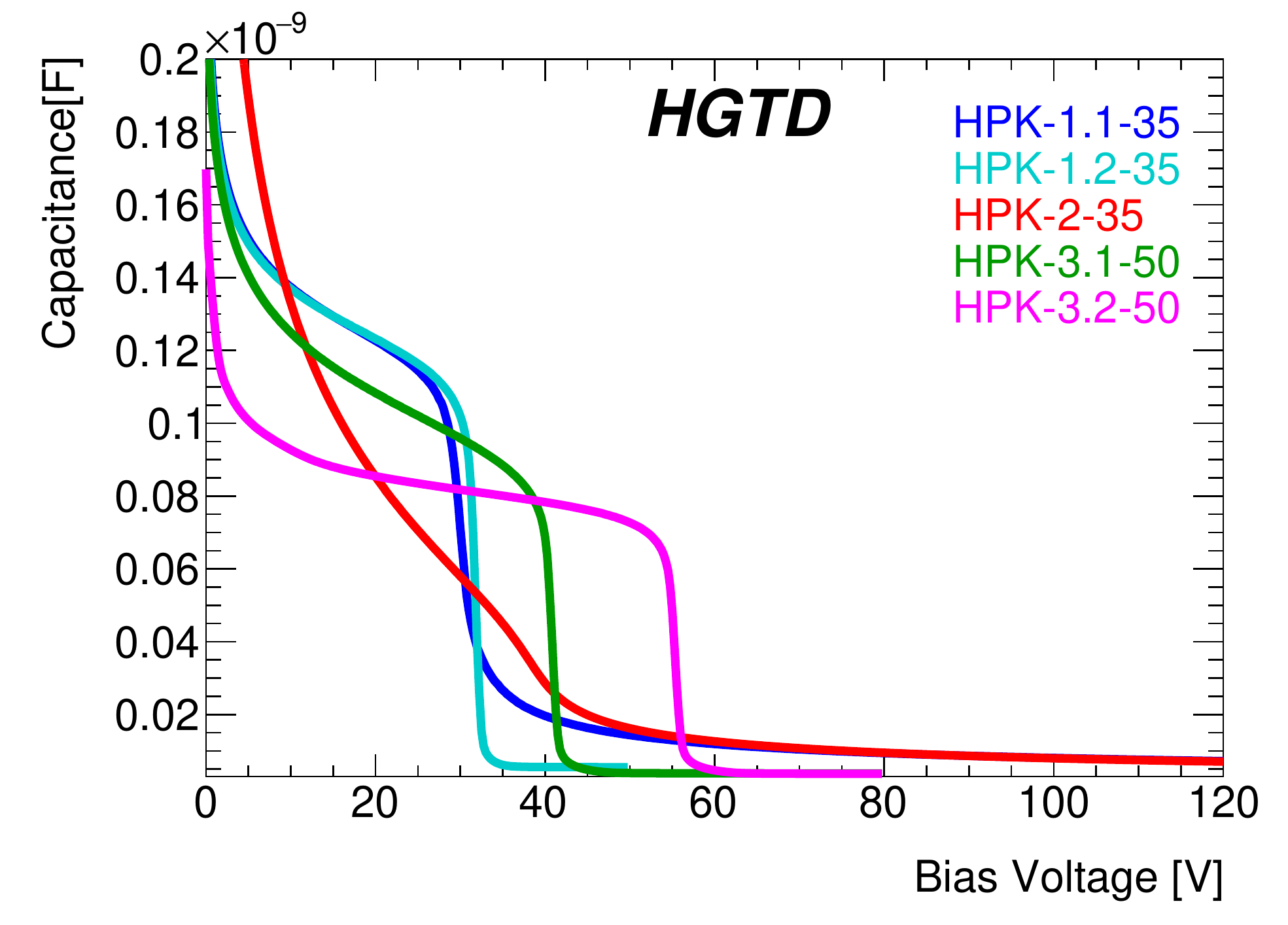}
\end{minipage}%
}%
\subfigure[]{
\begin{minipage}[t]{0.55\linewidth}
\includegraphics[width=1.0\linewidth]{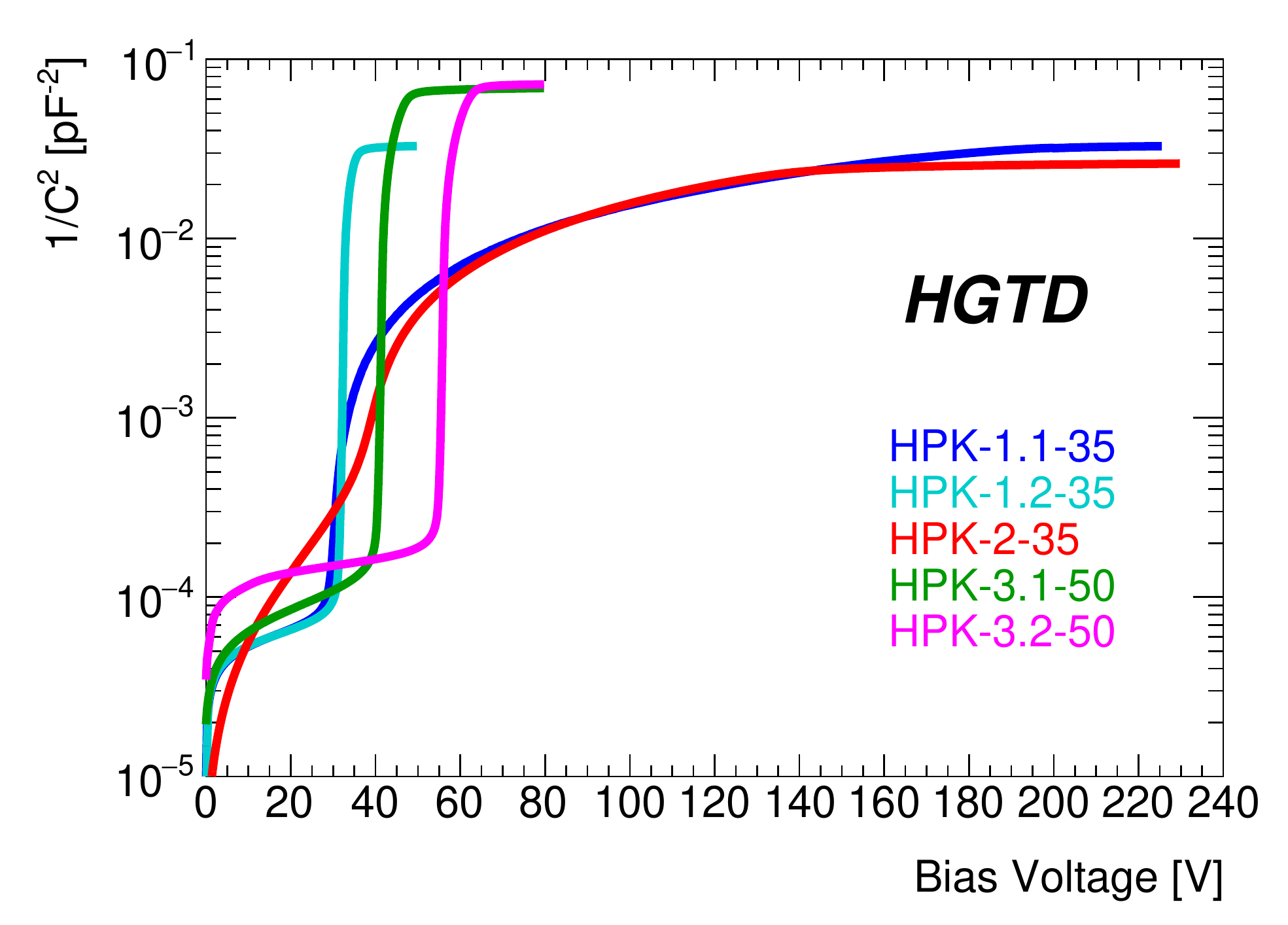}
\end{minipage}%
}%
\end{center}
\caption{C-V (a) and $\mathrm{1/C^2}$-V (b) curves of different types of HPK prototype sensor. The measurement is taken by single needle on the single pads with guard ring grounded.}
\label{Fig:4}
\end{figure}

To estimate the voltage required to deplete the gain layer (\VGL) and the bias to deplete the bulk (\VFD), C-V curves of each type's single pads are measured with probe needle and shown in Figure~\ref{Fig:4}(a).

For a typical p-n junction, the behavior of $\mathrm{1/C^2}$-V curve is closely related to the distribution of dopants \cite{rossi2006pixel}. It can be used to study the multiplication layer of the LGAD. Several parameters from $\mathrm{1/C^2}$-V curves in Figure~\ref{Fig:4}(b) are determined to compare the sensors. For each type of the sensor, the \VGL is estimated by finding the voltage where the $\mathrm{1/C^2}$-V curves start rising~\cite{HartmutHSTD12}, and the \VFD is the voltage where the curves saturate. The precise turning points are determined by extrapolating from the adjacent segments before and after the transition region and finding the intersections of the extrapolations. The gain layer depletion voltage is about \SI{31}{\volt} to \SI{33}{\volt} for HPK-1.1-35 and HPK-1.2-35, \SI{40}{\volt} to \SI{42}{\volt} for HPK-2-35 and HPK-3.1-50 and \SI{56}{\volt} for HPK-3.2-50.  

After gain layer depletion, the $\mathrm{1/C^2}$-V curves of HPK-1.2-35, HPK-3.1-50 and HPK-3.2-50 sensors rise much faster than other types and become flat at relatively low voltage, indicating a higher resistivity in the bulk. The full depletion voltages for HPK-1.2-35, HPK-3.1-50 and HPK-3.2-50 sensors are estimated to be in the range of 40-\SI{70}{\volt}, and those for HPK-1.1-35, HPK-2-35 are above \SI{100}{\volt}, indicting a lower resistive bulk. Detailed depletion voltages obtained from C-V are summarized in Table \ref{Tab:3}.
\begin{table}[htp]
\begin{center}
 \begin{tabular}{|c|c|c|}
    \hline
    Type&\VGL [\si{\volt}]& \VFD [\si{\volt}]\\
     \hline
    HPK-1.1-35&31 & 195\\
    \hline
    HPK-1.2-35&33&  36\\
    \hline
    HPK-2-35&40&    144\\
    \hline
    HPK-3.1-50&42& 49 \\
    \hline
    HPK-3.2-50&56&  64\\
     \hline
 \end{tabular}
 \end{center}
 \caption{Summary of \VGL and \VFD from single pad C-V measurements (Figure~\ref{Fig:4}).}
 \label{Tab:3}
\end{table} 
 
\subsection{Uniformity study of the gain layer}
\begin{figure}[htp]
\begin{center}
\subfigure[]{
\begin{minipage}[t]{0.55\linewidth}
\includegraphics[width=1\linewidth]{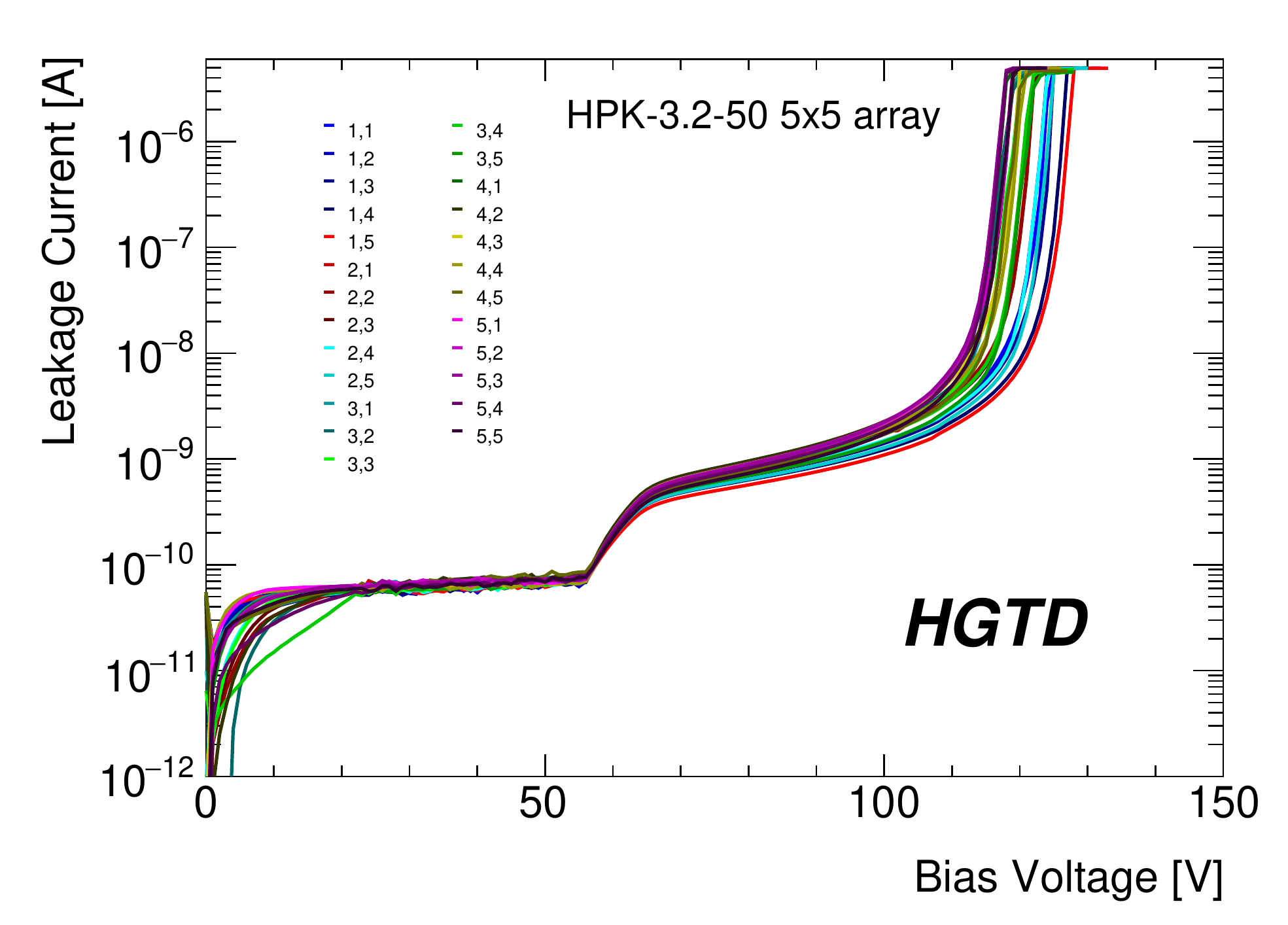}
\end{minipage}%
}%
\subfigure[]{
\begin{minipage}[t]{0.55\linewidth}
\includegraphics[width=1\linewidth]{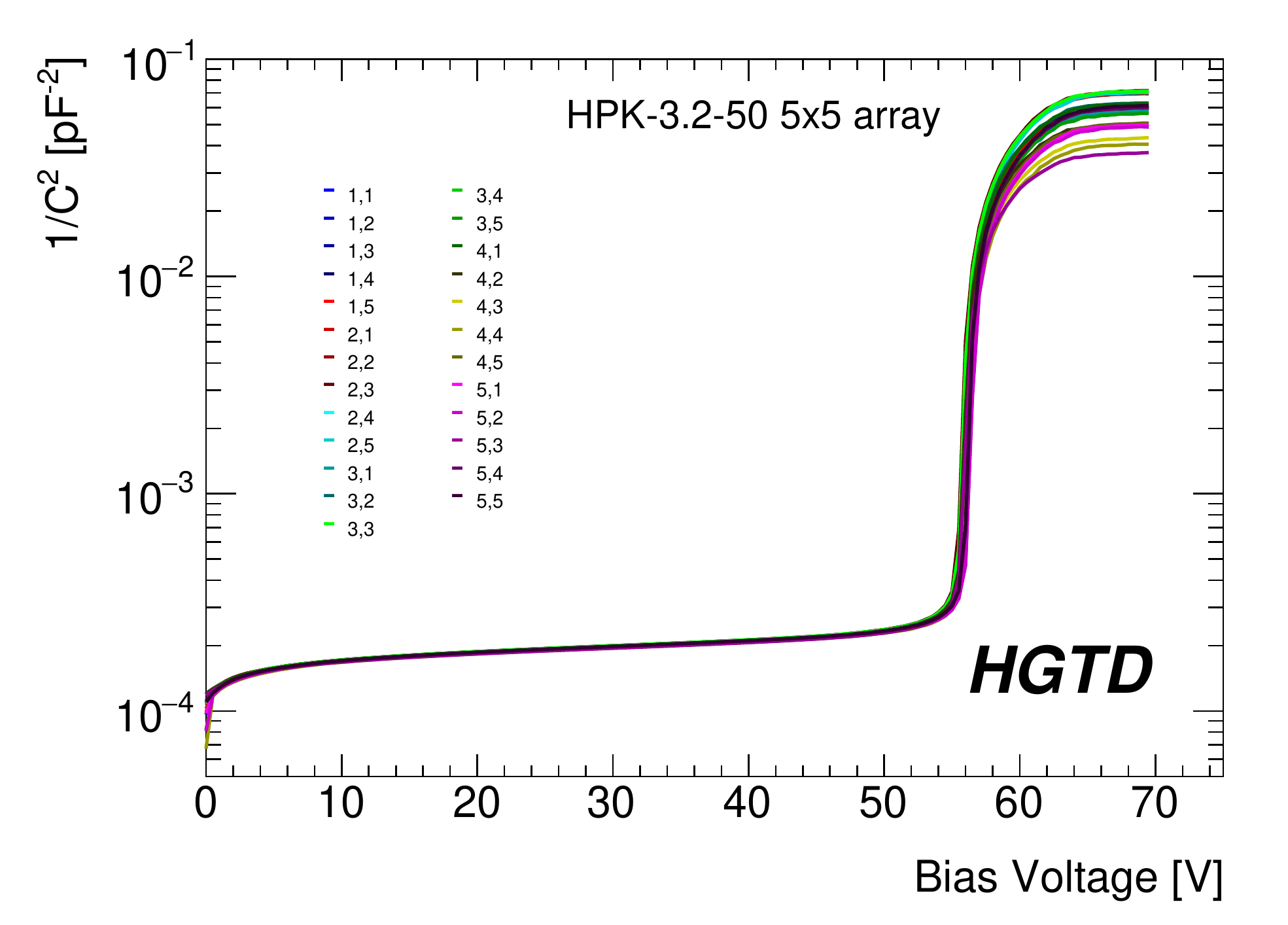}
\end{minipage}%
}%
\end{center}
\caption{I-V and C-V curves of an HPK-3.2-50 \num{5 x 5} used to measure the \VGL variation at the foot region (near \SI{56}{V}). The measurement is taken by a dedicate probe card with guard ring and neighbor pads grounded.}
\label{Fig:10}
\end{figure}

\begin{figure}[htp]
\begin{center}
\subfigure[]{
\begin{minipage}[t]{0.55\linewidth}
\includegraphics[width=1\linewidth]{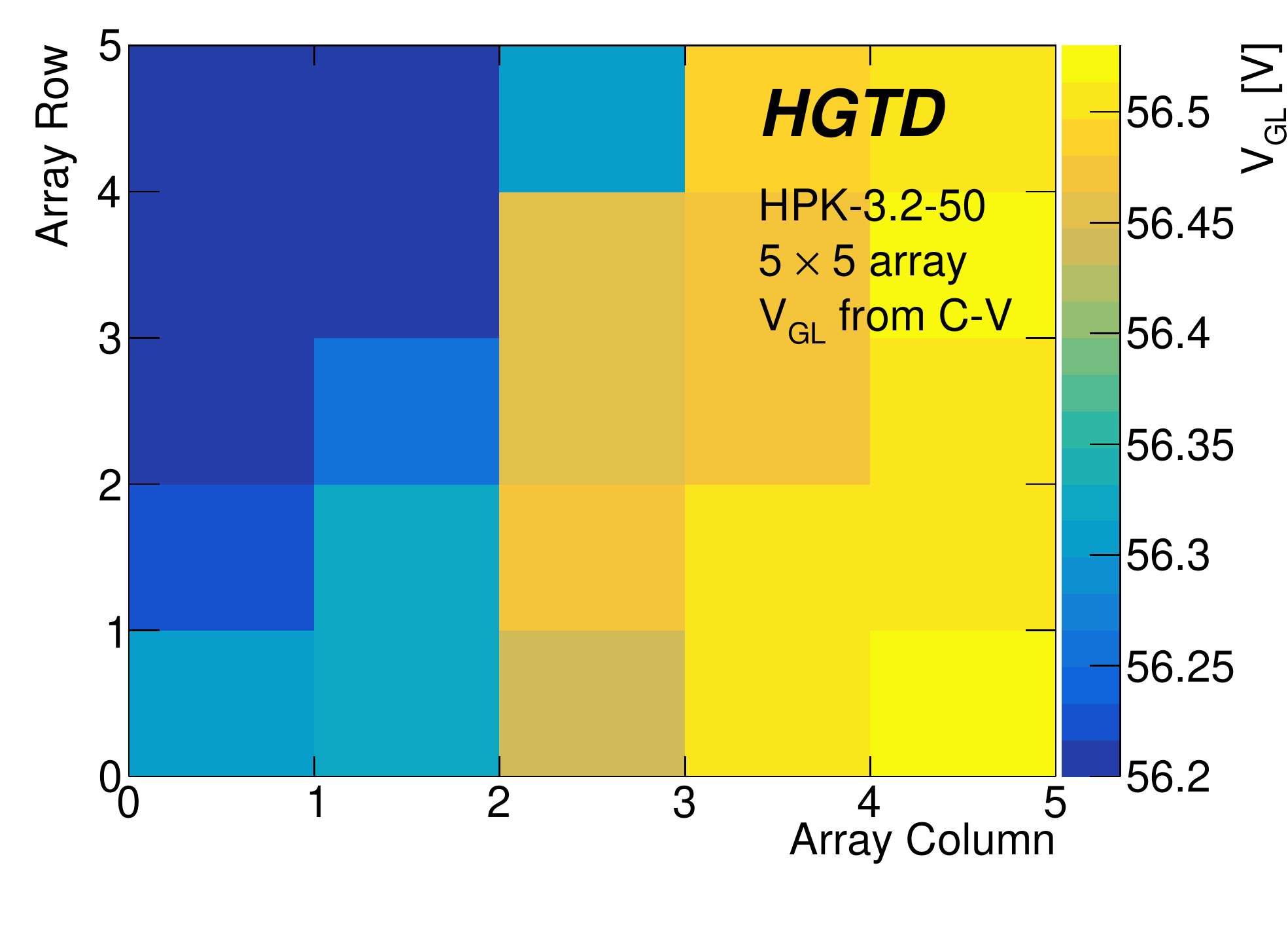}
\end{minipage}%
}%
\subfigure[]{
\begin{minipage}[t]{0.55\linewidth}
\includegraphics[width=1\linewidth]{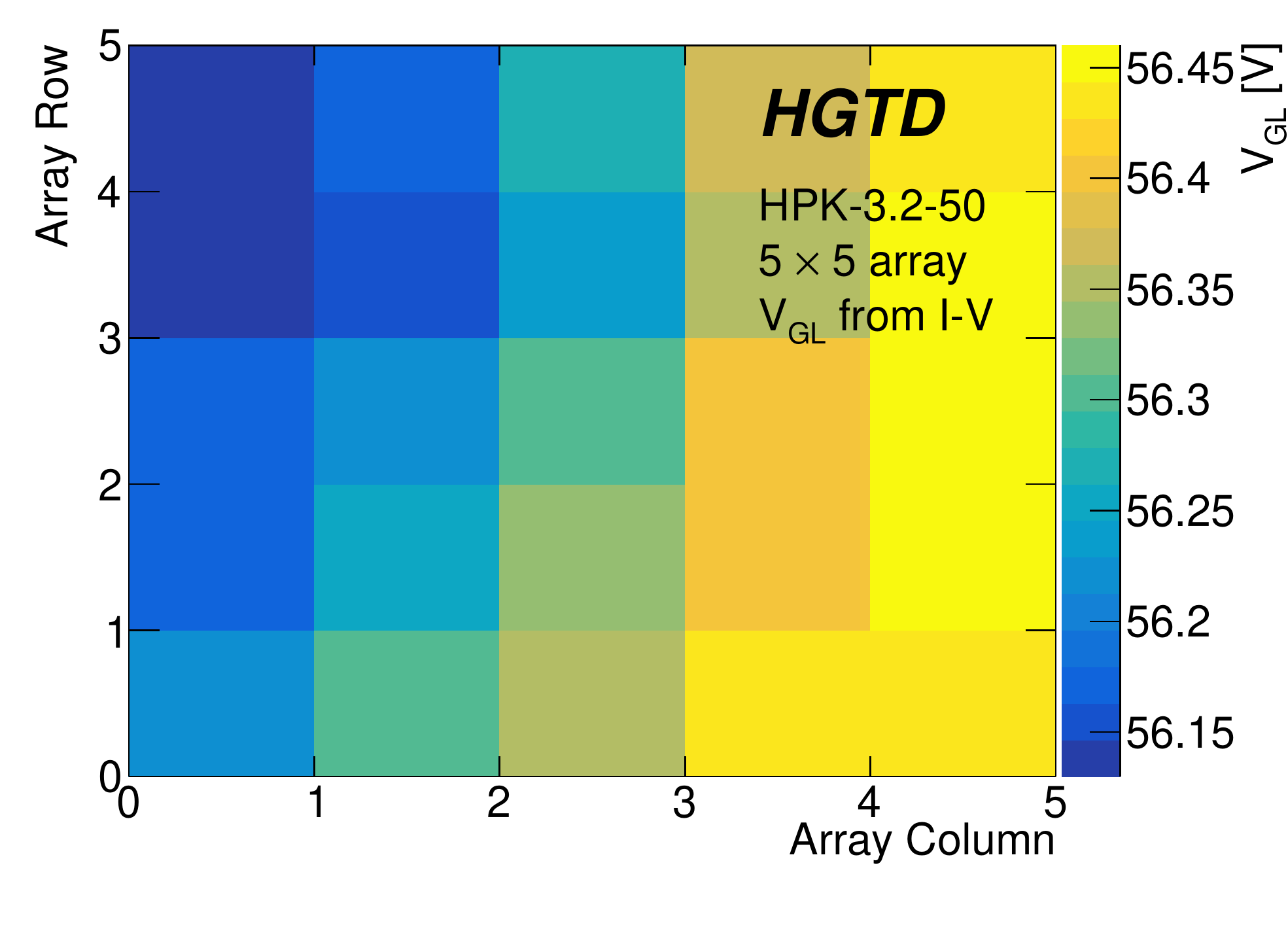}
\end{minipage}%
}%
\vskip\baselineskip

\subfigure[]{
\begin{minipage}[t]{0.55\linewidth}
\includegraphics[width=1\linewidth]{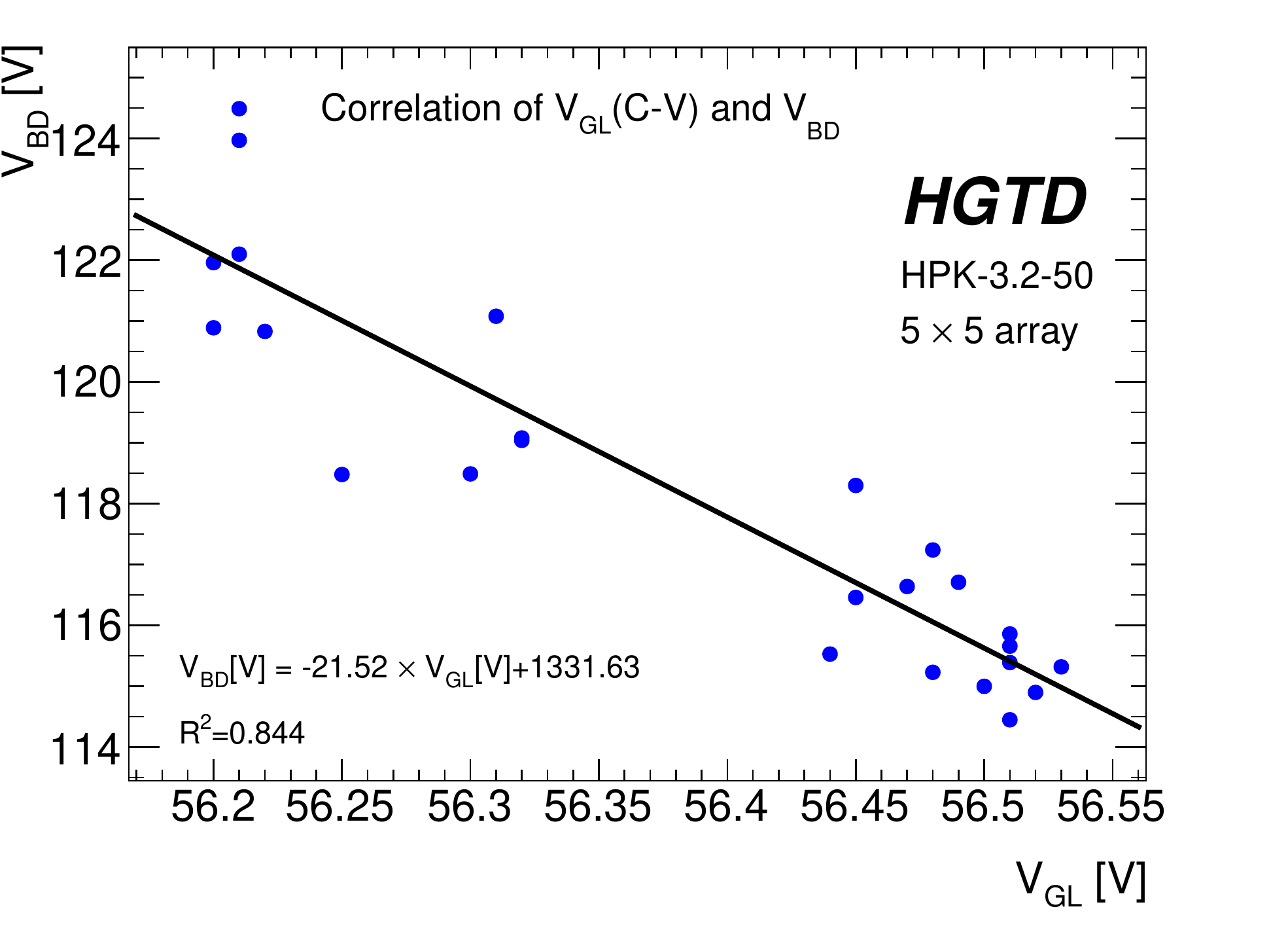}
\end{minipage}%
}%
\subfigure[]{
\begin{minipage}[t]{0.55\linewidth}
\includegraphics[width=1\linewidth]{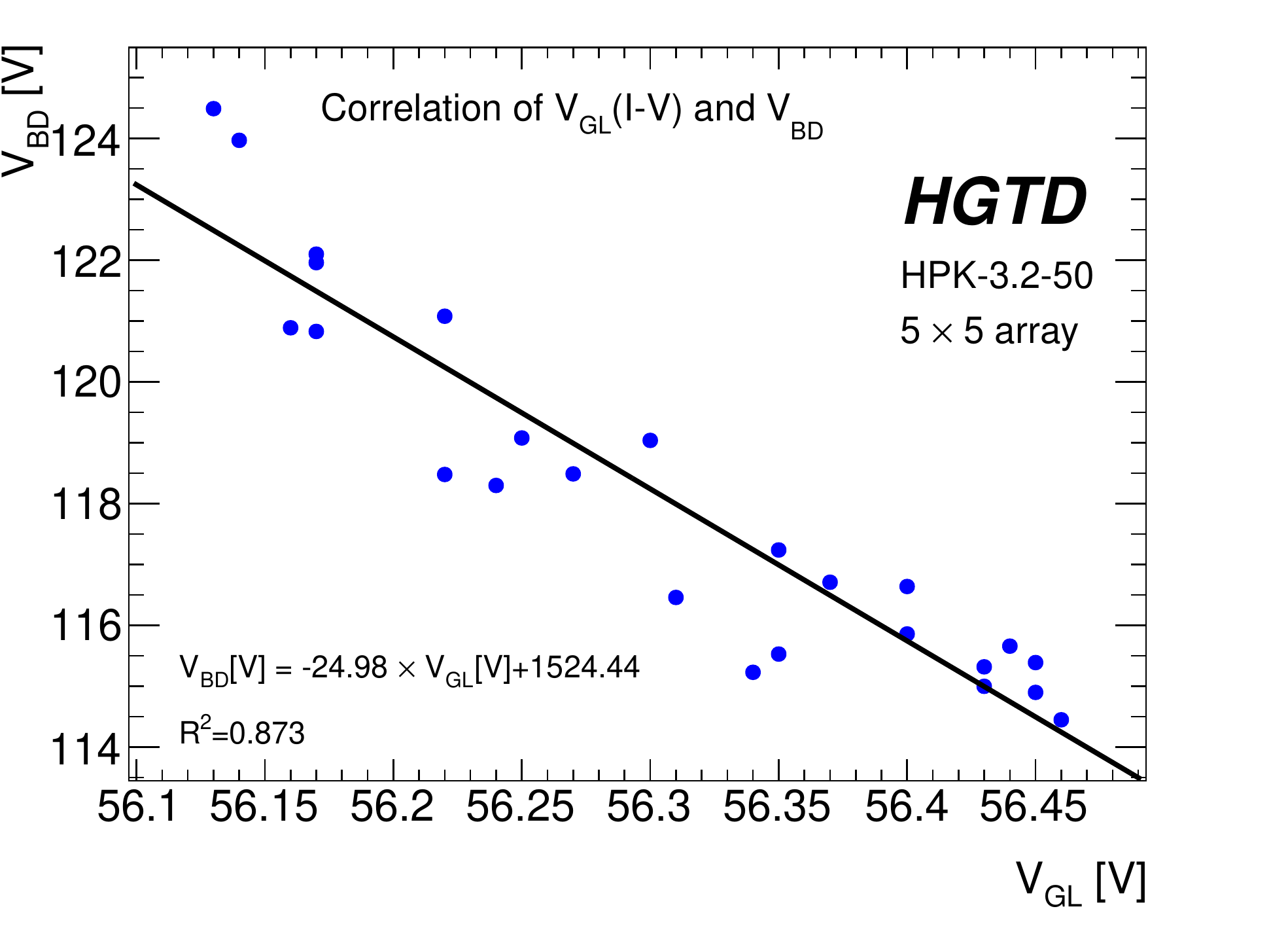}
\end{minipage}%
}%
\end{center}
\caption{Distribution of \VGL from C-V (a) and I-V (b) on an HPK-3.2-50 \num{5 x 5} pads array. The correlation between each of them and \VBD are shown in the bottom. Linear fits are applied for each correlation and the fitted function as well as $\mathrm{R^2}$ (squared correlation coefficient) are shown in the legend. The analyzed data are from Figure~\ref{Fig:10}.}
\label{Fig:8}
\end{figure}

Since the front-end chip used to used to read out the sensor provides a common ground to the sensor and the bias voltage is common to a full \num{30 x 15} pad array simultaneously, the uniformity of gain layer should be well controlled. Due to the fact that a small variation on boron implantation dose would result in a large variation on \VBD as shown in Figure \ref{Fig:5}. Assuming a fixed acceptor distribution, the total boron implantation dose should be proportional to the \VGL. Thus a measurement of \VGL variation is important. The resistance of a biased sensor is dependent on the depth of the depletion zone \cite{pellegrini2014technology}. Thus I-V curves (Figure \ref{Fig:10}(a)) can also be used to estimate \VGL besides the C-V curves (Figure \ref{Fig:10}(b)). Here, the I-V curves of a \num{5 x 5} array are measured with finer steps, slower HV ramping rate to derive the \VGL of each pad. Figure \ref{Fig:8} (a) and (b) show the distribution of \VGL from I-V and C-V on a same HPK-3.2-50 \num{5 x 5} array. The correlation between \VGL and \VBD is shown in Figure \ref{Fig:8} (c) and (d). The mean value as well as variation are summarized in the Table \ref{Tab:4}. The results from both methods are close. Finally, a 0.2\% variation on boron implantation dose is observed, which results in a 2.5\% variation on \VBD.

\begin{table}[htp]
\begin{center}
 \begin{tabular}{|c|c|c|c|}
    \hline
    Method & \VGL from C-V & \VGL from I-V & \VBD \\
     \hline
    Mean value [V] &56.38 & 56.31 & 118.12 \\
    \hline
    RMS/Mean $\times$ 100\%  &0.22\%  &  0.19\%&   2.48\% \\
    \hline
 \end{tabular}
 \end{center}
 \caption{Comparison of different methods to characterize the gain layer uniformity in an HPK-3.2-50 \num{5 x 5} array.}
 \label{Tab:4}
\end{table} 

\section{Inter-pad Gap}
\label{S:5}

\begin{figure}[htp]
\begin{center}
\subfigure[]{
\begin{minipage}[t]{0.55\linewidth}
\includegraphics[width=1\linewidth]{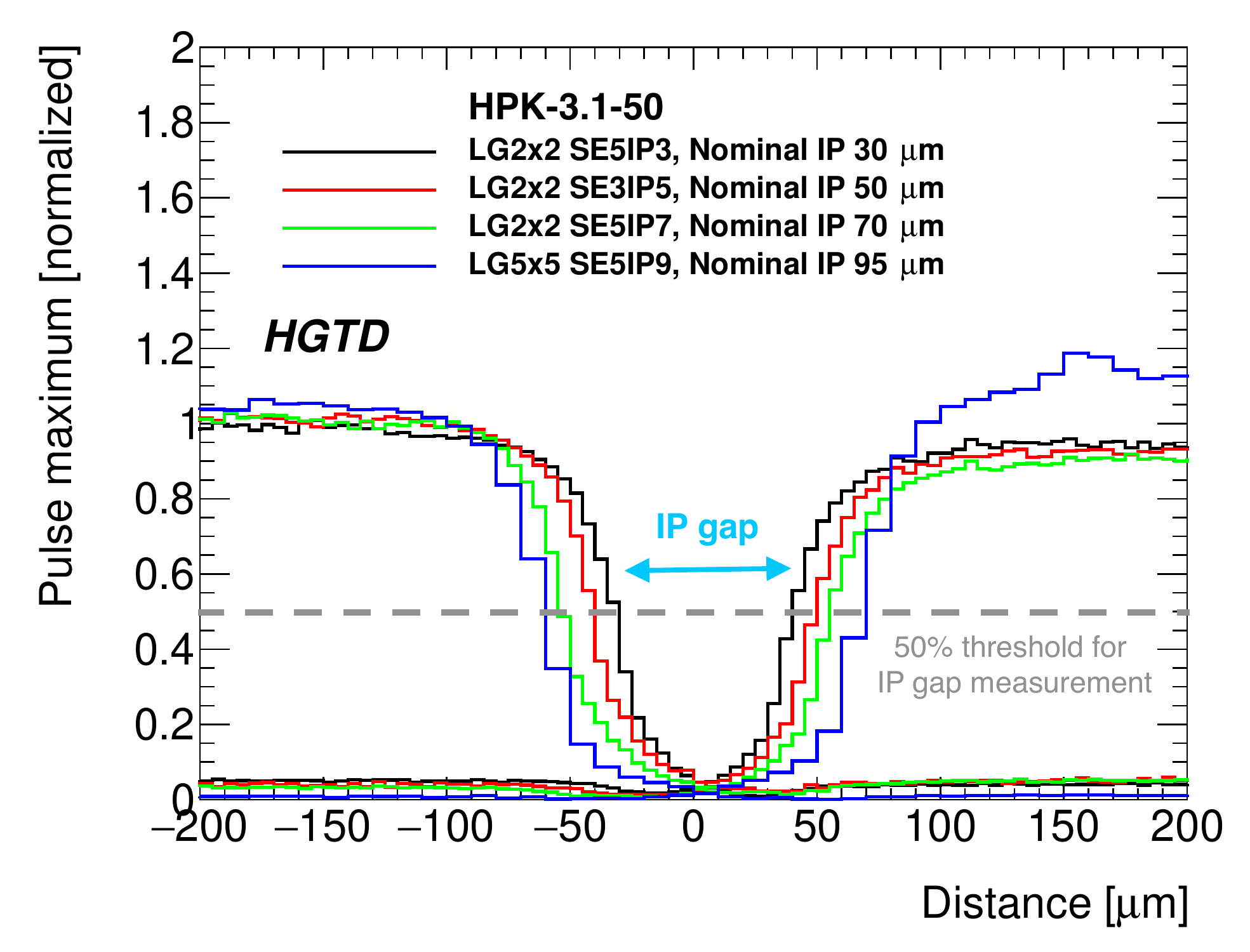}
\end{minipage}%
}%
\subfigure[]{
\begin{minipage}[t]{0.55\linewidth}
\includegraphics[width=1\linewidth]{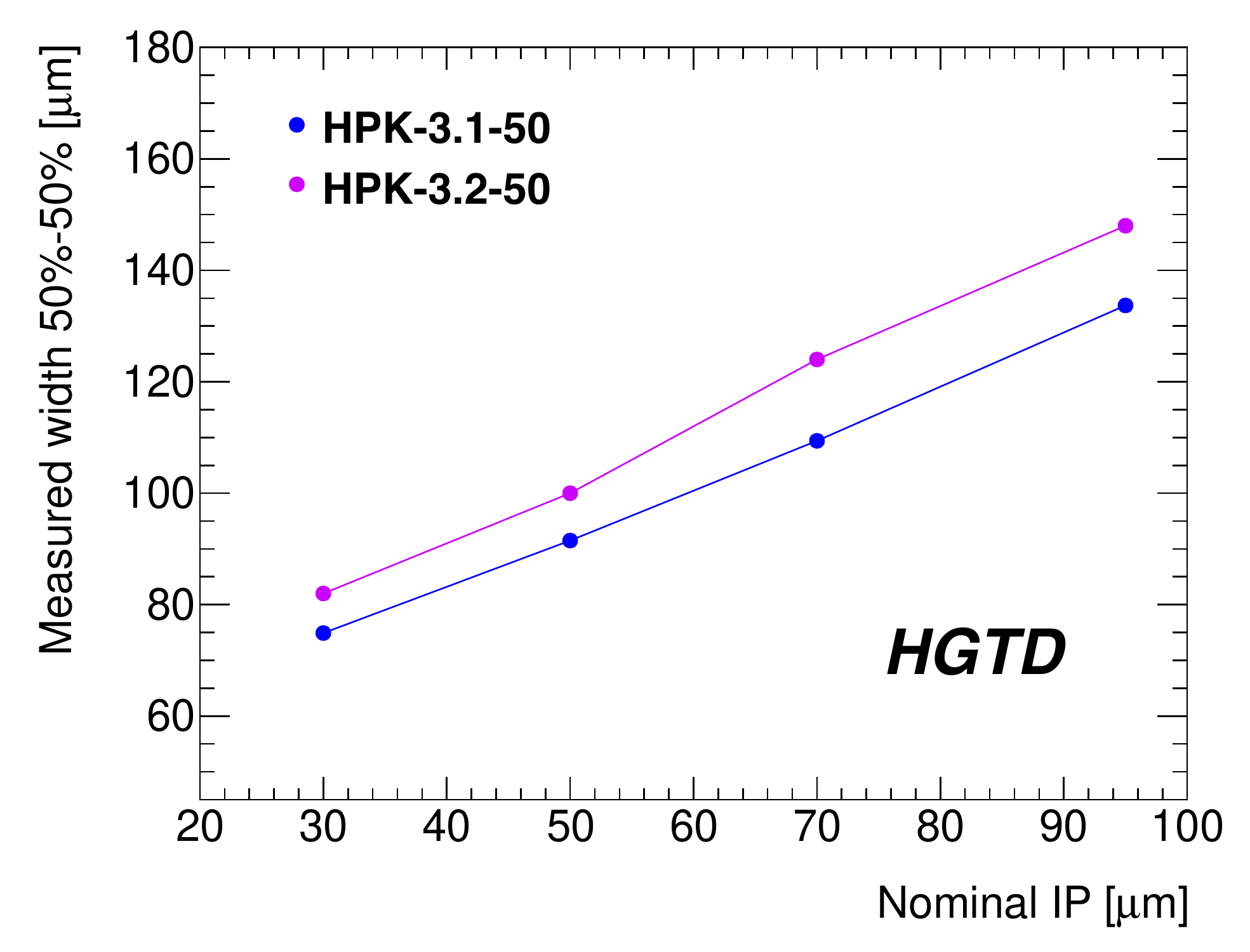}
\end{minipage}%
}%
\end{center}
\caption{(a) IP gap measurement of HPK-3.1-50 \num{2 x 2} arrays by laser scan~\cite{HGTDPublicPlots}. (b) Correlations between nominal and measured IP gap distances of HPK sensors~\cite{mschLGAD}.}
\label{Fig:6}
\end{figure}

The \SI{1060}{\nano\metre} infrared laser TCT is used for IP gap studies by scanning over the region between two adjacent pads. The nominal value of the IP gap is defined as the distance between the multiplication region of adjacent pads, which ranges from \SI{30}{\micro\metre} to \SI{95}{\micro\metre} for HPK \num{2 x 2} LGAD arrays. Experimentally, the IP gap is here defined by the region between adjacent pads where the pulse height is below 50\% of that in the central area of a pad. As shown in Figure~\ref{Fig:6}, the measured IP gap is always about \SI{40}{\micro\metre} to \SI{50}{\micro\metre} larger than the nominal due to that the field lines under the multiplication area are bent towards the peripheral region where JTE~(Junction Termination Extension) lies with no gain. This effect would be reduced by irradiation because of removed JTE doping and increased bulk gain. The IP gap of HPK-3.1-50 is always measured to be slightly lower than the one of HPK-3.2-50 and an IP gap of \SI{70}{\micro\metre} is achieved at the lowest nominal IP gap of \SI{30}{\micro\metre}, which corresponds to a fill factor around 90\%.

\section{Time resolution and collected charge}

\begin{figure}[htp]
\begin{center}
\subfigure[]{
\begin{minipage}[t]{0.55\linewidth}
\includegraphics[width=1\linewidth]{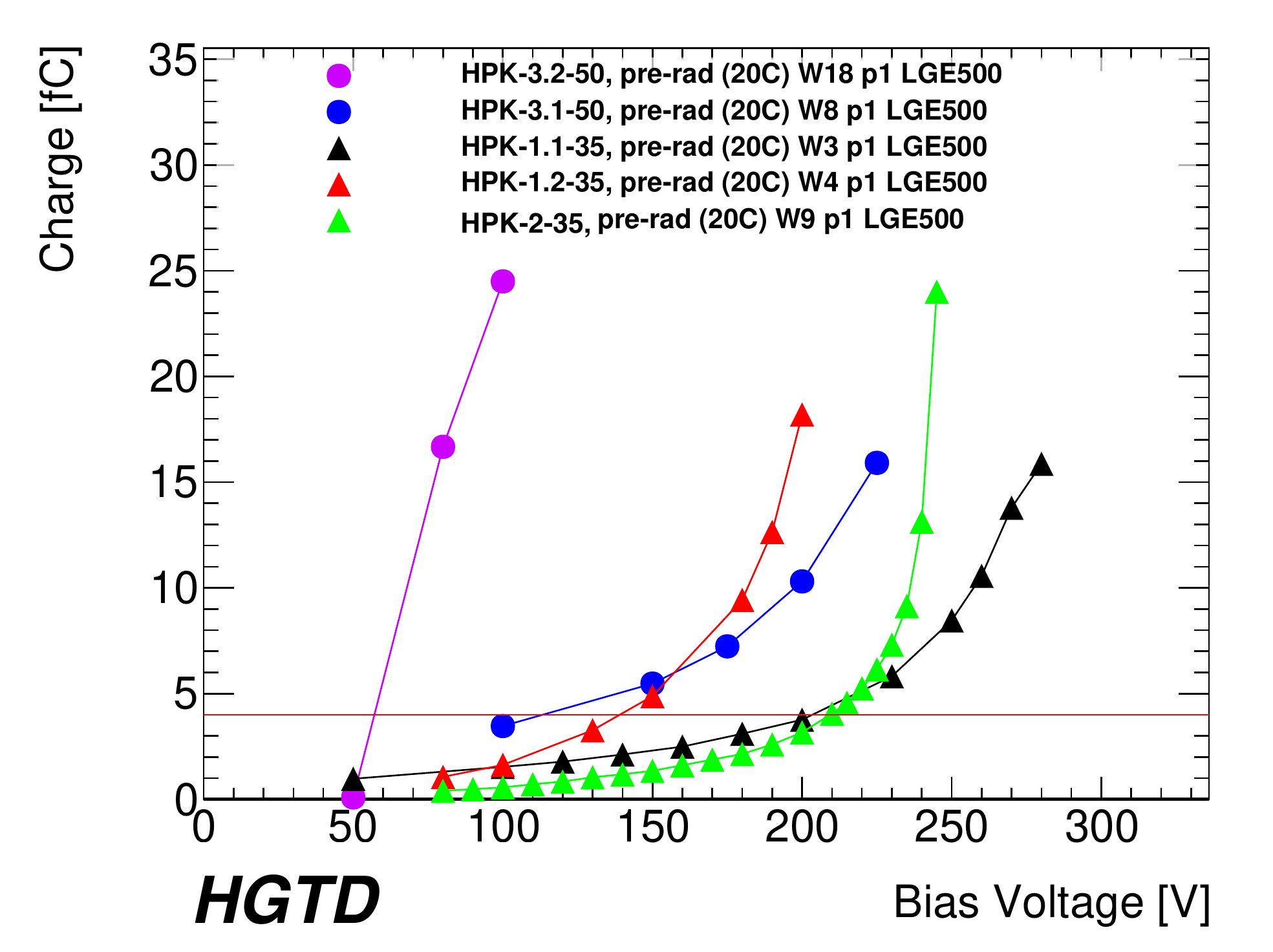}
\end{minipage}%
}%
\subfigure[]{
\begin{minipage}[t]{0.55\linewidth}
\includegraphics[width=1\linewidth]{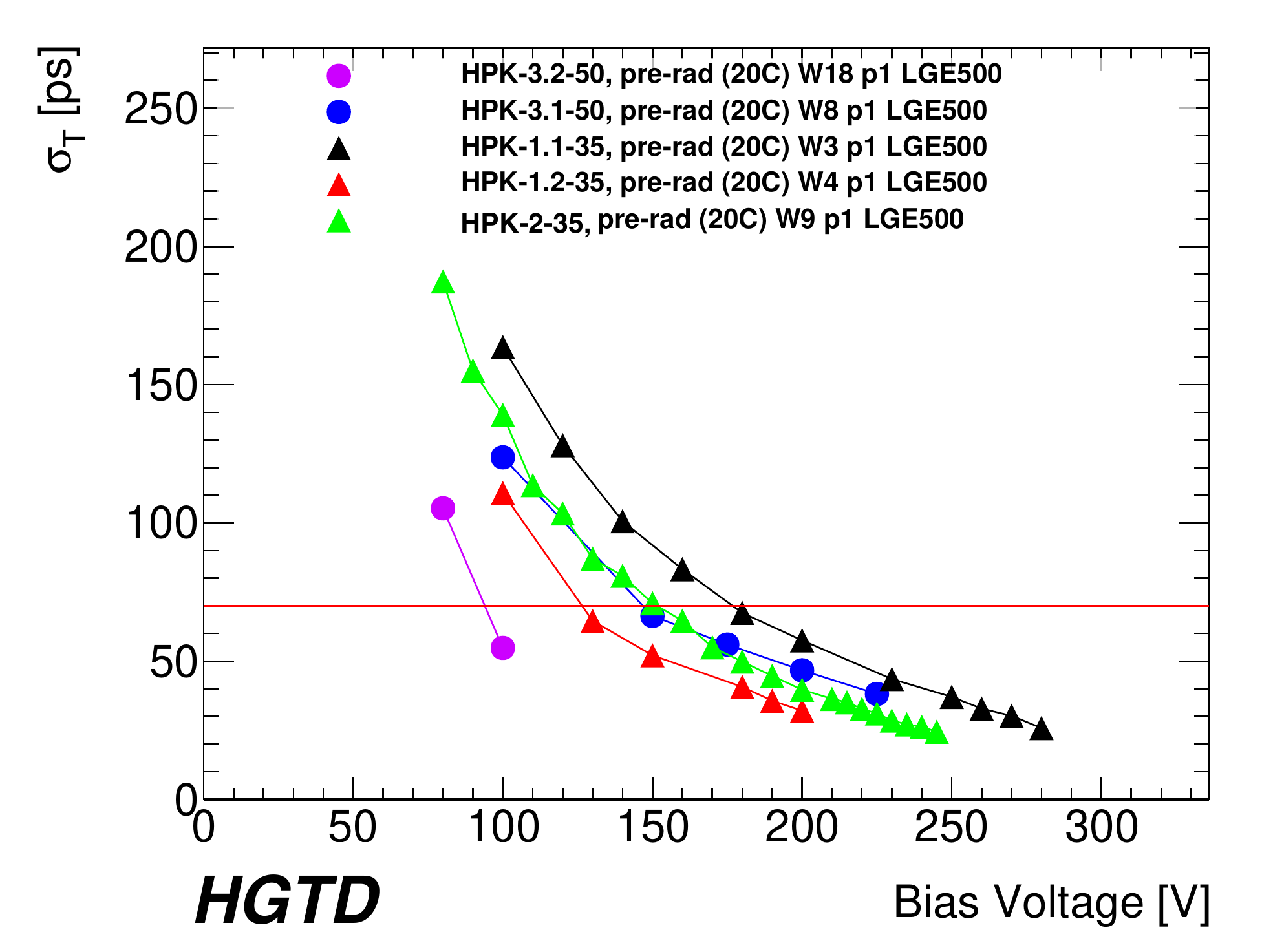}
\end{minipage}%
}%
\end{center}
\caption{Collected charge (a) and time resolution (b) of different types of HPK prototype sensor before irradiation. The requirements of the HGTD project are indicated by the horizontal red lines.}
\label{Fig:7}
\end{figure}

The collected charge as well as the time resolution at different bias voltages are measured with a $\mathrm{^{90}Sr}$ $\beta$-scope system \cite{GALLOWAY201919}. The time resolution are extracted with a reference fast HPK LGAD with \SI{15}{\pico\second} of time resolution which also acts as a trigger. The result are shown in Figure \ref{Fig:7}(a), from which we can see different performance resulting from different design.
Due to high implantation and early breakdown, HPK-3.2-50 has to be operated at low bias, just after full depletion, and the carrier velocity is not saturated. This results in a reduced time resolution performance.
The time resolution under different bias voltages are summarized in \ref{Fig:7}(b), which shows that all types except HPK-3.2-50 have reached the time resolution below \SI{30}{\pico\second} before breakdown. The sensors are irradiated by proton from accelerators (CYRIC, Los Alamos) and neutrons from a reactor source (JSI), the performances after irradiation are summarized in another paper \cite{Xintalk}.

\section{Conclusion}
\label{S:7}
LGAD sensors produced by HPK have been studied extensively by the ATLAS HGTD group. The breakdown voltage and depletion voltage of each type have been characterized by I-V and C-V measurements to determine the window for detector operation. A measured effective inter-pad gap of \SI{70}{\micro\metre} has been achieved by HPK-3.1-50 sensors. The variation of \VGL is measured to characterize the fluctuation of gain layer implantation and a 0.2\% variation is observed.

The fractions of good pads and perfect sensors are estimated for each array size. Good uniformities are observed in leakage currents and breakdown voltages and the feasibility of large size sensors production has been demonstrated. Time resolution and collected charges are measured by a $\beta$-scope and time resolution below \SI{30}{\pico\second} is achieved.

\section*{Acknowledgement}
This work was supported by the United States Department of Energy, grant DE-FG02-04ER41286, “the Fundamental Research Funds for the Central Universities” of China (grant WK2030040100), the National Natural Science Foundation of China (No. 11961141014), the State Key Laboratory of Particle Detection and Electronics (SKLPDE-ZZ-202001), the Hundred Talent Program of the Chinese Academy of Sciences (Y6291150K2), the CAS Center for Excellence in Particle Physics (CCEPP), the MINECO, Spanish Government, under grant RTI2018-094906-B-C21, the Slovenian Research Agency (project J1-1699 and program P1-0135), the U.S.A. Department of Energy under grant contact DE-SC0012704 and partially carried out at the USTC Center for Micro and Nanoscale Research and Fabrication and partially performed within the CERN RD50 collaboration. The contributions from UCSC technical staff and students is acknowledged.






\bibliographystyle{elsarticle-num}
\bibliography{sample}







\end{document}